\documentclass[showpacs,preprintnumbers,amsmath,amssymb]{revtex4}
\usepackage{mathrsfs}
\usepackage{graphicx}

\begin{document}

\title{Statefinder Parameters for
Interacting Phantom Energy with Dark Matter}

\author{Baorong Chang\footnote{changbaorong@student.dlut.edu.cn},
Hongya Liu\footnote{Corresponding author: hyliu@dlut.edu.cn},
Lixin Xu, Chengwu Zhang and Yongli Ping}

\address{School of Physics \& Optoelectronic Technology, Dalian
University of Technology, Dalian, 116024, P. R. China}

\keywords{statefinder parameter, scaling solution, dark energy.}
\pacs{98.80.-k, 98.80.Es}

\begin{abstract}
We apply in this paper the statefinder parameters to the interacting
phantom energy with dark matter. There are two kinds of scaling
solutions in this model. It is found that the evolving trajectories
of these two scaling solutions in the statefinder parameter plane
are quite different, and that are also different from the
statefinder diagnostic of other dark energy models.

\end{abstract}

\maketitle



Cosmic observations indicate that our universe is undergoing an
accelerated expansion and the dominated component of the present
universe is dark energy \cite{xu1,xu2,xu3,zhang2,zhang3}. The
Wilkinson Microwave Anisotropy Probe (WMAP) satellite experiment
tells us that dark energy, dark matter, and the usual baryonic
matter occupy about $73\%$, $23\%$, and $4\%$ of the total energy of
the universe, respectively. The accelerated expansion of the present
universe is attributed to the dark energy whose essence is quite
unusual and there is no justification for assuming that it resembles
known forms of matter or energy. Candidates for dark energy have
been widely studied and focus on the cosmological constant $\Lambda$
\cite{sahni9904398,cosmological constant} with $\mathcal {W}=-1$, a
dynamically evolving scalar field (quintessence)
\cite{quintessence1,quintessence2} with $\mathcal {W}>-1$ and
phantom \cite{phantom} with $\mathcal {W}<-1$. Recently, a study of
high-$\mathcal {Z}$ ($\mathcal {Z}$ is redshift) SNe Ia \cite{SNe}
find that the equation of state of dark energy has a $99\%$
probability of being $\mathcal {W}<-1$ if no priors are placed on
$\Omega_{m}^{0}$. When these SNe results are combined with CMB and
2dFGRS the $95\%$ confidence limits on an unevolving equation of
state are $-1.46<\mathcal {W}<-0.78$ \cite{xu3,SNe} which is
consistent with estimates made by other groups \cite{zhang2,zhang3}.
In order to obtain $\mathcal {W}<-1$, phantom field with reverse
sign in its dynamical term may be a simplest way and can be regarded
as one of interesting possibilities describing dark energy
\cite{Caldwell}. However, the other physical properties of phantom
energy are rather weird, as they include violation of the
dominant-energy condition, naive superluminal sound speed and
increasing energy density with time. The latter property ultimately
leads to unwanted future singularity called big rip which had been
considered in \cite{9912054}. This singularity is characterized by
the divergence of the scale factor in a finite time in future
\cite{sigularity}.

Many authors have discussed various kinds of phantom field models to
avoid the cosmic doomsday \cite{avoid big rip}, which require a
special class of phantom field potentials with a local maximum.
Moreover, the energy density of the phantom field increases with
time, while the energy density of the matter fluid decreases as the
universe expands. Why are the energy density of dark matter and the
phantom energy density of the same order just at the present epoch?
This coincidence problem becomes more difficult to solve in the
phantom model without the suitable interaction \cite{without
interaction}. But Guo et al. in Ref. \cite{0411524,0412624} proposed
a suitable interaction in the phantom field model, and the
coincidence problem can be avoided. Moreover in Ref. \cite{0412624},
the universe also avoids the big rip. In Ref. \cite{0412624},
considering a universe model which contains phantom field $\phi$ and
the dark matter $\rho_{dm}$. The Friedmann equation in a spatially
flat FRW metric can be written as follows:
\begin{equation}
H^{2}=\frac{\kappa^{2}}{3}(\rho_\phi+\rho_{dm}),\label{H^2}
\end{equation}
and
\begin{equation}
\dot{H}=-\frac{\kappa^{2}}{2}(\rho_{dm}-{\dot{\phi}}^{2}),\label{dotH}
\end{equation}
where $\kappa^{2}\equiv8\pi{G}$, $\rho_{dm}$ is the energy density
of the dark matter, and the dark matter possesses the equation of
state $P_{dm}=0$. The energy density and pressure of the phantom
field $\phi$ are $\rho_{\phi}$ and $P_{\phi}$, respectively,
\begin{equation}
\rho_\phi=-\frac{1}{2}{\dot{\phi}}^{2}+V(\phi),\label{density}
\end{equation}
\begin{equation}
P_\phi=-\frac{1}{2}{\dot{\phi}}^{2}-V(\phi).\label{pressure}
\end{equation}
where $V(\phi)$ is the phantom field potential,
$V(\phi)=V_{0}\exp(-\lambda\kappa\phi)$. Here, we postulate that
the two component $\rho_{\phi}$ and $\rho_{dm}$, interact through
the interaction term $Q$ according to
\begin{equation}
\dot{\rho}_\phi+3H(\rho_\phi+P_\phi)=-Q,\label{DE}
\end{equation}
\begin{equation}
\dot{\rho}_{dm}+3H(\rho_{dm}+P_{dm})=Q.\label{DM}
\end{equation}
Here, the interaction term has the specific formation
\begin{equation}
Q=3cH(\rho_{\phi}+\rho_{dm}).\label{Q}
\end{equation}
where $c$ is a dimensionless parameter denoting the transfer
strength. We define the following dimensionless variables
\begin{equation}
x=\frac{\kappa\dot{\phi}}{\sqrt{6}H},\qquad\qquad
y=\frac{\kappa\sqrt{V}}{\sqrt{3}H},\qquad\qquad
z=\frac{\kappa\sqrt{\rho_{dm}}}{\sqrt{3}H}.\label{def}
\end{equation}
Thus the fractional densities of $\rho_\phi$ and $\rho_{dm}$ are
\begin{equation}
\Omega_\phi=-x^{2}+y^{2},\qquad\qquad \Omega_{dm}=z^{2}.
\end{equation}
The evolution equations (\ref{DE}) and (\ref{DM}) can be written
as the following set of equations:
\begin{eqnarray}
x^{'}&=&-3x(1+x^{2}-\frac{1}{2}z^{2}-\frac{1}{2}cx^{-2})-\frac{\sqrt{6}}{2}\lambda{y^{2}},\label{newx}\\
y^{'}&=&-3y(x^{2}+\frac{\sqrt{6}}{6}\lambda{x}-\frac{1}{2}z^{2}),\label{newy}\\
z^{'}&=&-3z(\frac{1}{2}+x^{2}-\frac{1}{2}z^{2}-\frac{1}{2}cz^{-2}),\label{newz}
\end{eqnarray}
where the prime denotes a derivative with respect to the logarithm
of the scale factor, $N=\ln{a}$. From the definitions of these new
variables, we find that the equation of state of phantom $\mathcal
{W}_{\phi}$ is $\mathcal {W}_{\phi}=(-x^2-y^2)/(-x^2+y^2)$. So we
get the effective parameter of equation-of-state of the phantom is
\begin{equation}
\mathcal {W}_\phi^{eff}=\frac{-x^{2}-y^{2}+c}{-x^2+y^2}.
\end{equation}
and the effective equation of state for the total cosmic fluid is
\begin{equation}
\mathcal {W}_{eff}=-x^{2}-y^{2}.
\end{equation}
Guo et al. in Ref. \cite{0412624} had concluded that in the case of
the interaction (\ref{Q}), there exist two stable scaling solutions,
the climbing-up scaling solution and the rolling-down scaling
solution. In this model the universe evolves from a matter-dominated
phase to a scaling solution, which is characterized by a constant
ratio of the energy densities of the dark matter and the phantom
field, which may give a phenomenological solution of the coincidence
problem. In the climbing-up case, the phantom field initially climbs
up, the effective equation of state $\mathcal {W}_{eff}$ realizes a
transition from $\mathcal {W}_{eff}>-1$ to $\mathcal {W}_{eff}<-1$,
and the universe ends with a big rip. In the rolling-down case, the
phantom field initially rolls down, the effective equation of state
tends to above $-1$ and realizes a transition from $\mathcal
{W}_{eff}<-1$ to $\mathcal {W}_{eff}>-1$, in this case the cosmic
doomsday is avoided and the universe accelerates forever. There have
been so many models that proposed to explain the cosmic acceleration
and solve, or at least alleviate, the coincidence problem
\cite{quintessence2,coincidence problem,grqc0311067}. In order to
discriminate this interacting phantom model from others, and
differentiate these two scaling solutions further, we refer to a
cosmological diagnostic pair $\{r,s\}$ called statefinder which is
introduced by Sahni et al. in \cite{statefinder} and defined as
\begin{equation}
r\equiv\frac{\dddot{a}}{aH^{3}},\qquad
s\equiv\frac{r-1}{3(q-1/2)},
\end{equation}
Here $q$ is the deceleration parameter. The statefinder is a
"geometrical" diagnostic in the sense that it depends on the
expansion factor and hence on the metric describing space-time.
Since different cosmological models involving dark energy exhibit
qualitatively different evolution trajectories in the $s-r$ plane,
this statefinder diagnostic can differentiate various kinds of dark
energy models. For the spatially flat LCDM cosmological model, the
statefinder parameters correspond to a fixed point $\{r=1,s=0\}$. S
far, some models including the cosmological constant, quintessence,
phantom, quintom, the Chaplygin gas, braneworld models, holographic
models, interacting and coupling dark energy models
\cite{grqc0311067,statefinder,models for statefinder}, have been
successfully differentiated. For example, the quintessence model
with inverse power law potential, the phantom model with power law
potential and the Chaplygin gas model all tend to approach the LCDM
fixed point, but for quintessence and phantom models the
trajectories lie in the regions $s>0, r<1$ while for Chaplygin gas
model the trajectories lie in the regions $s<0, r>1$. In this paper
we apply the statefinder diagnostic to the coupled phantom model. To
begin with, we use another form of statefinder parameters in terms
of the total energy density $\rho$ and the total pressure $p$ in the
universe:
\begin{equation}
r=1+\frac{9(\rho+p)}{2\rho}\frac{\dot{p}}{\dot{\rho}},\qquad
s=\frac{(\rho+p)}{p}\frac{\dot{p}}{\dot{\rho}}.
\end{equation}
Since the total energy of the universe is conserved, we have
$\dot{\rho}=-3H(\rho+p)$. Then making use of
$\dot{\rho_\phi}=-3H(1+\mathcal {W}_{\phi}^{eff})\rho_\phi$, we can
get
\begin{eqnarray}
r&=&1-\frac{3}{2}\mathcal {W}_{\phi}^{'}\Omega_\phi+\frac{9}{2}\mathcal {W}_\phi(1+\mathcal {W}_{\phi}^{eff})\Omega_\phi,\\
s&=&1-\frac{\mathcal {W}_{\phi}^{'}}{3\mathcal {W}_\phi}+\mathcal
{W}_{\phi}^{eff},
\end{eqnarray}
where $\mathcal {W}_{\phi}^{'}=\frac{d\mathcal {W}_{\phi}}{dN}$. and
the deceleration parameter is also given
\begin{equation}
q=\frac{1}{2}(1+3\mathcal {W}_{\phi}\Omega_\phi)
\end{equation}

\begin{figure}
\begin{center}
\includegraphics[width=2.4in]{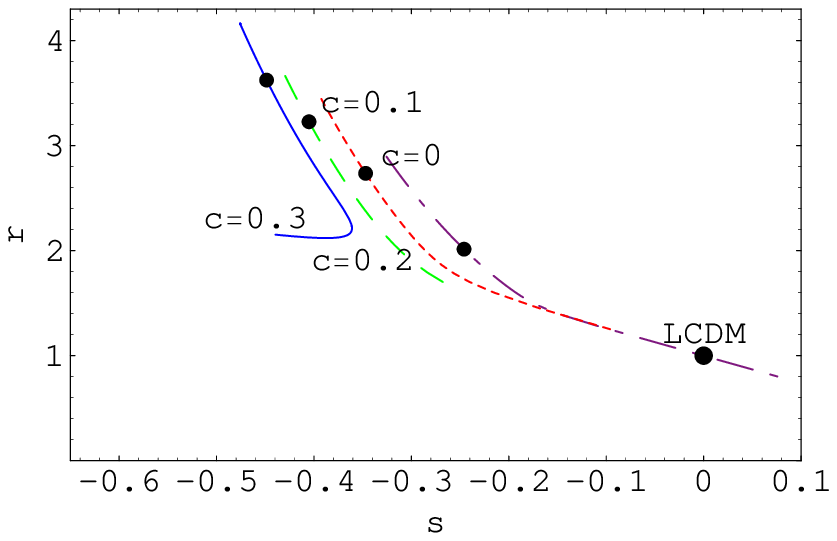}\includegraphics[width=2.4in]{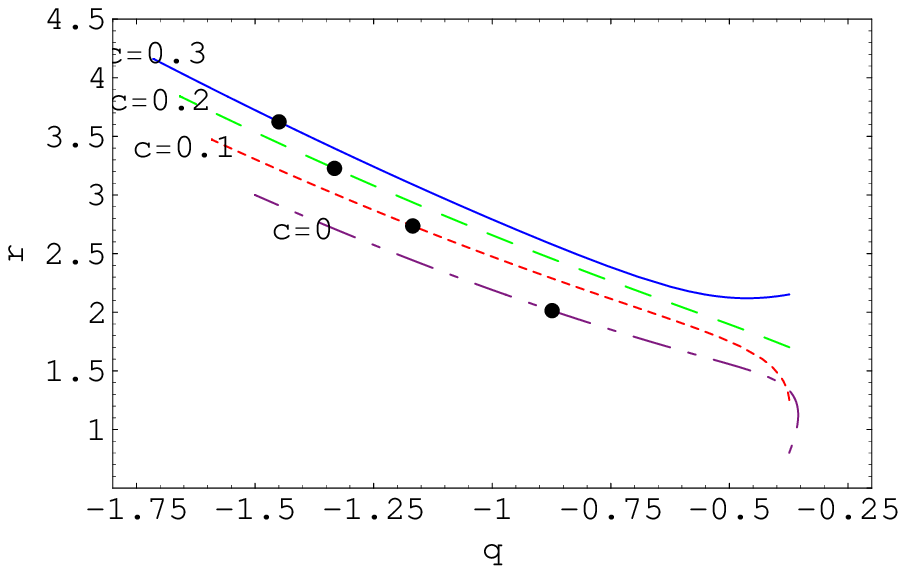}
\end{center}
\caption{The left figure is $s-r$ diagram of the climbing-up
scaling solution with $x_0<0$. The right figure is $q-r$ diagram
of the climbing-up scaling solution with $x_0<0$. The curves
evolve in the variable interval $N\in[-2,20]$. Selected curves for
$\lambda=1$, $c=0$, $0.1$, $0.2$ and $0.3$ respectively. Dots
locate the current values of the statefinder
parameters.}\label{sr1}
\end{figure}

\begin{figure}
\begin{center}
\includegraphics[width=2.4in]{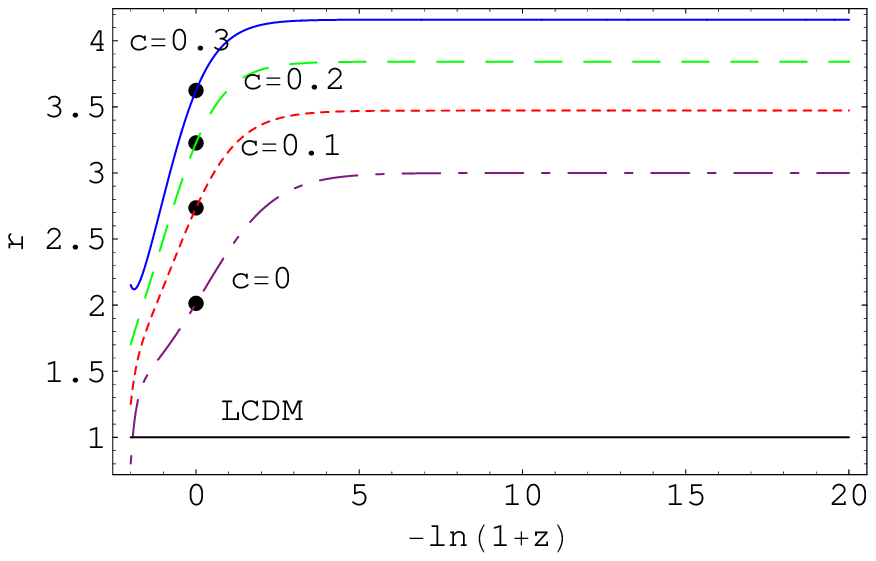}\includegraphics[width=2.4in]{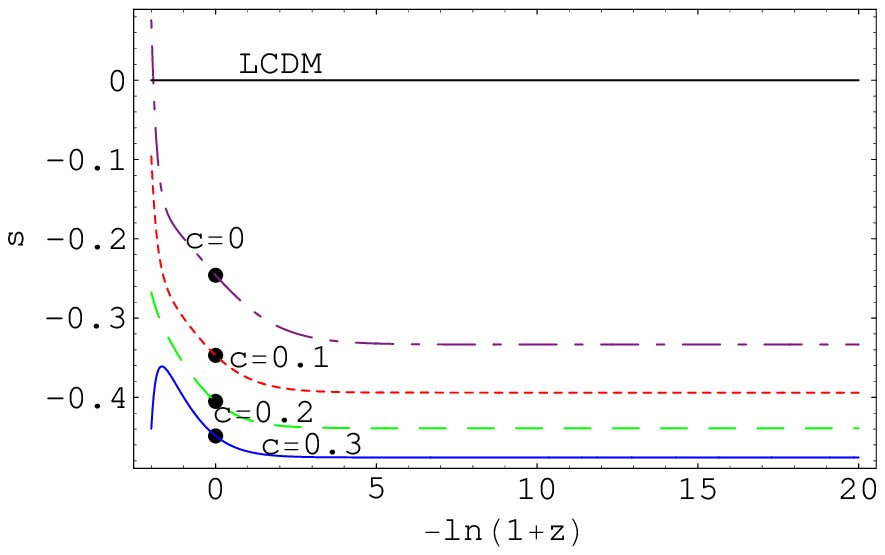}\includegraphics[width=2.4in]{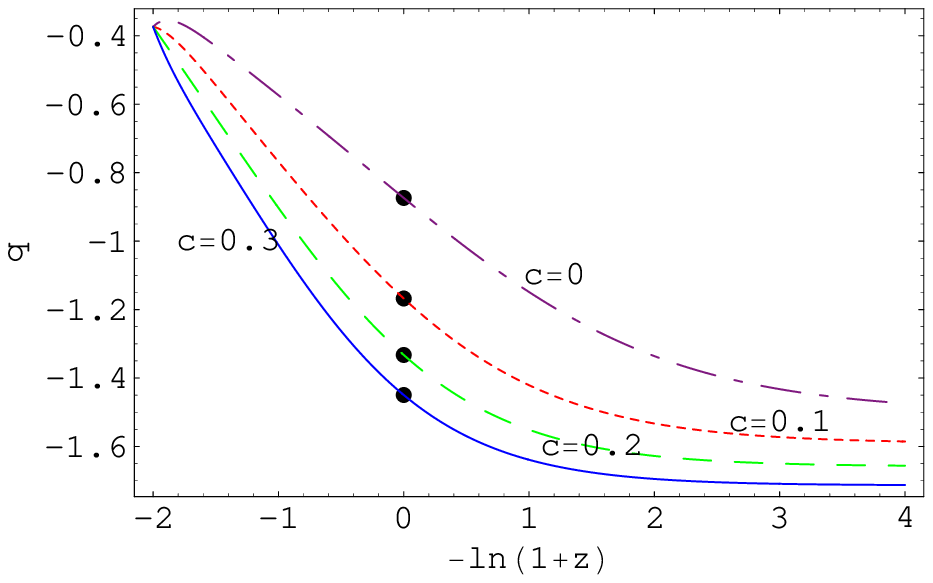}
\end{center}
\caption{The figures are the statefinder parameters versus
redshift diagram of the climbing-up scaling solution with $x_0<0$.
The left figure is $-\ln(1+z)$ --- $r$ diagram in the variable
interval $N\in[-2,20]$. The middle figure is $-\ln(1+z)$ --- $s$
diagram in the variable interval $N\in[-2,20]$. The right figure
is $-\ln(1+z)$ --- $q$ diagram in the variable interval
$N\in[-2,4]$. Selected curves for $\lambda=1$, $c=0$, $0.1$, $0.2$
and $0.3$ respectively. Dots locate the current values of the
statefinder parameters.}\label{nr1}
\end{figure}

In the following we will discuss the statefinder for two scaling
solutions with different conditions in Ref \cite{0412624}, and
investigate how the interaction between phantom and dark matter
influences the evolution of the universe. Firstly, we discuss the
scaling solution $B$ in the case of the interaction form (\ref{Q}),
which exists for $x_B<0$ ($x_B$ is the initially velocity of the
phantom field) and
$0<{c}\leq{f(\frac{-\lambda-\sqrt{\lambda^2+12}}{2\sqrt{6}})}$
($f(x)\equiv{x(2x+\sqrt{6}\lambda/3)(1-\sqrt{6}x\lambda/3)}$ is
defined as a cubic function) and corresponds to a climbing-up
phantom field. In this scenario, the universe ends with a big rip.
In Fig. \ref{sr1} we show the time evolution of the statefinder
pairs $\{r,s\}$ and $\{r,q\}$ for the climbing-up scaling solution.
The plot is for variable interval $N\in[-2,20]$, and the selected
evolution trajectories of $r(s)$ and $r(q)$ correspond to
$\lambda=1$, $c=0$, $c=0.1$, $0.2$ and $0.3$ respectively. $c=0$
represents no interaction between phantom field and dark matter,
$c=0.1$, $0.2$ and $0.3$ represent the different transfer strength
of interaction between DE and DM. We see clearly that the curve will
pass through the LCDM fixed point when there is no coupling ($c=0$)
between phantom field and dark matter, while the distant from the
curves to LCDM scenario is considerable when exists the interaction
between phantom field and dark matter. The statefinder pair
$\{r,s\}$ lies in the regions $s<0$, $r>1$, which is different from
other quintessence and phantom model. In Fig. \ref{nr1} we plot the
evolution trajectories of the statefinder parameters versus redshift
diagram of the climbing-up scaling solution with $x_0<0$. In the
plots of the $-ln(1+z)$ --- $r$ and $-ln(1+z)$ --- $s$, we can see
clearly that the interaction between phantom field and dark matter
causes the big deviation between statefinder parameters and the LCDM
scenario.

\begin{figure}
\begin{center}
\includegraphics[width=2.4in]{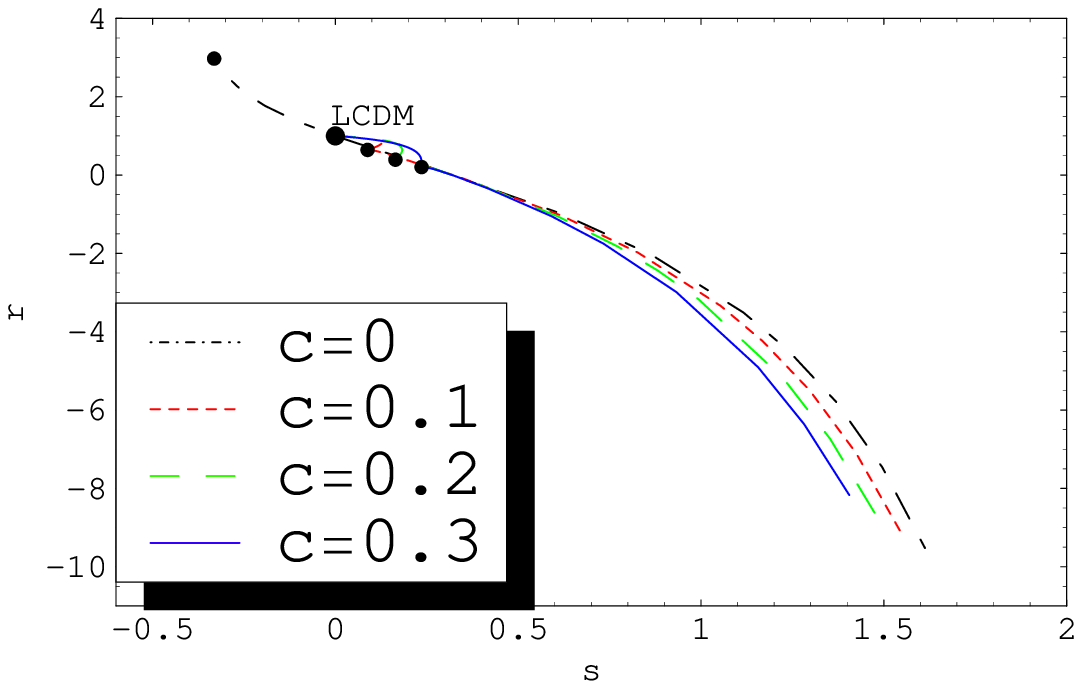}\includegraphics[width=2.4in]{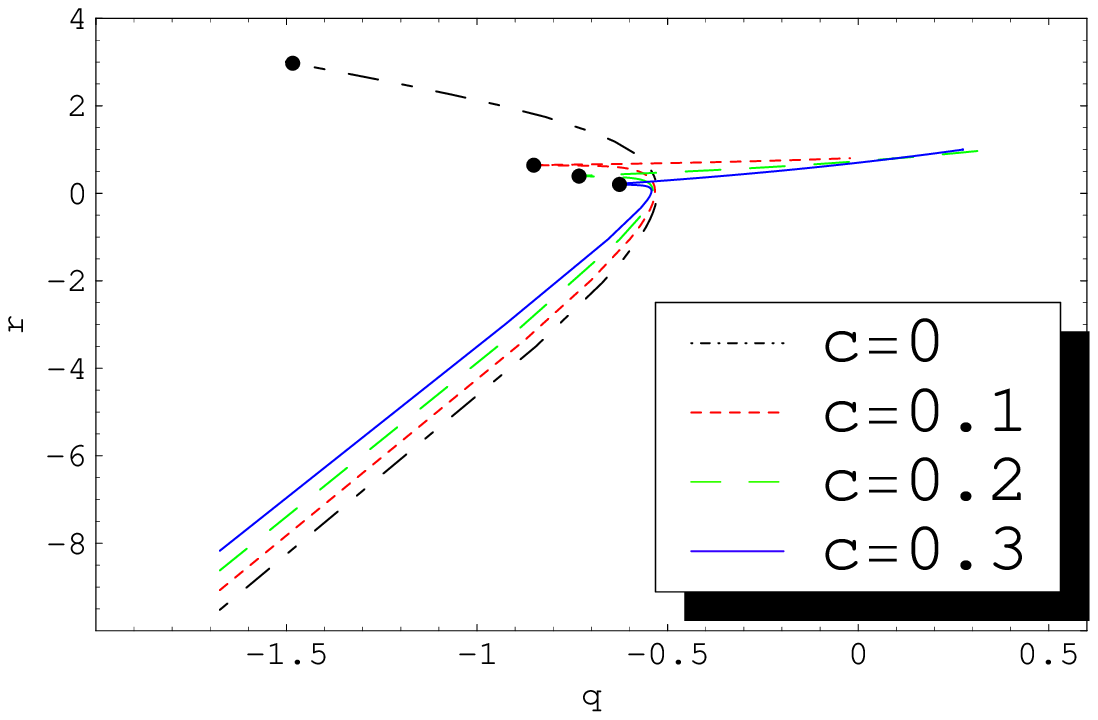}
\end{center}
\caption{The left figure is $s-r$ diagram of the rolling-down
scaling solution with $x_0>0$. The right figure is $q-r$ diagram
of the rolling-down scaling solution with $x_0>0$. The curves
evolve in the variable interval $N\in[-2,20]$. Selected curves for
$\lambda=1$, $c=0$, $0.1$, $0.2$ and $0.3$ respectively. Dots
locate the current values of the statefinder
parameters.}\label{sr2}
\end{figure}

\begin{figure}
\begin{center}
\includegraphics[width=2.4in]{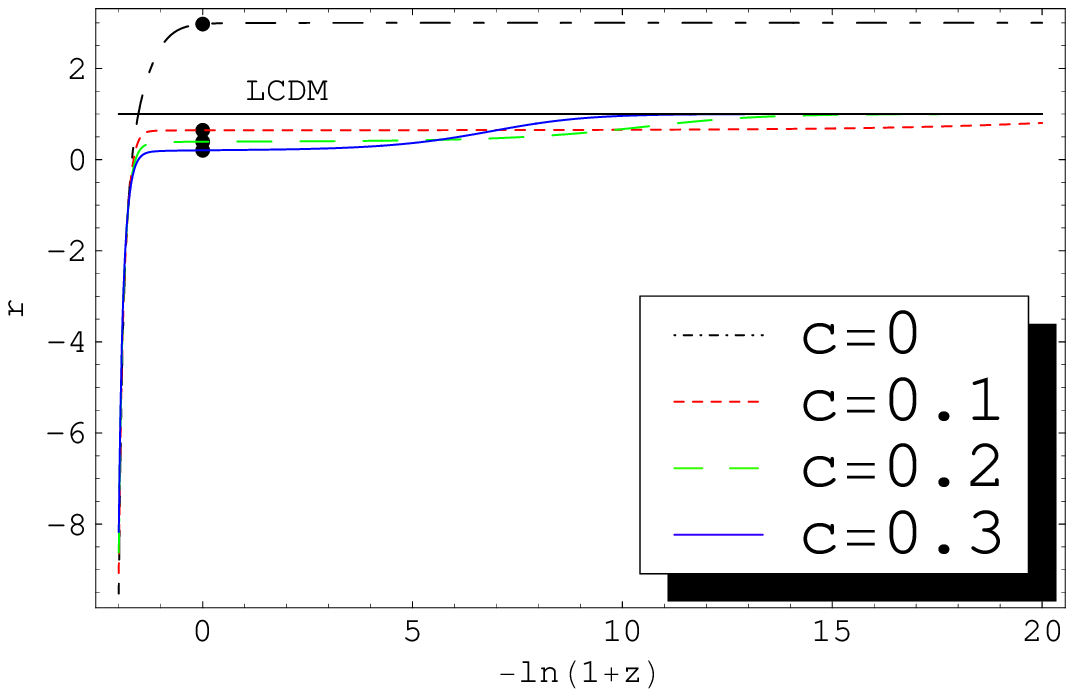}\includegraphics[width=2.4in]{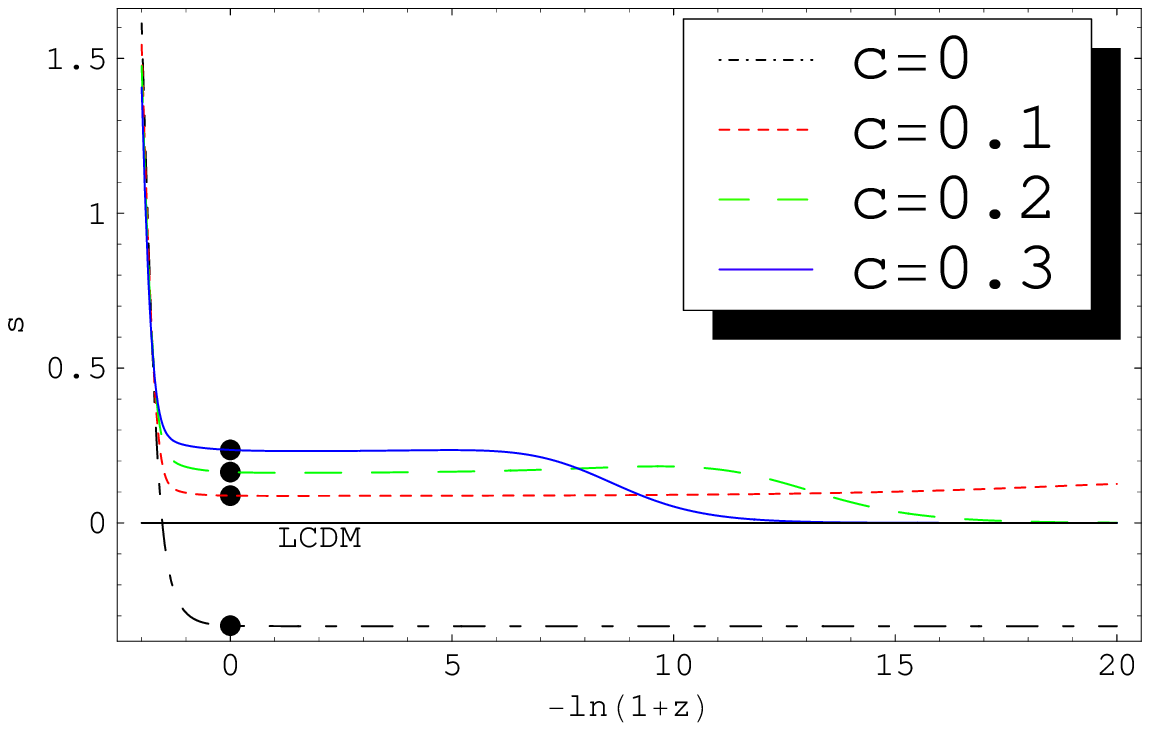}\includegraphics[width=2.4in]{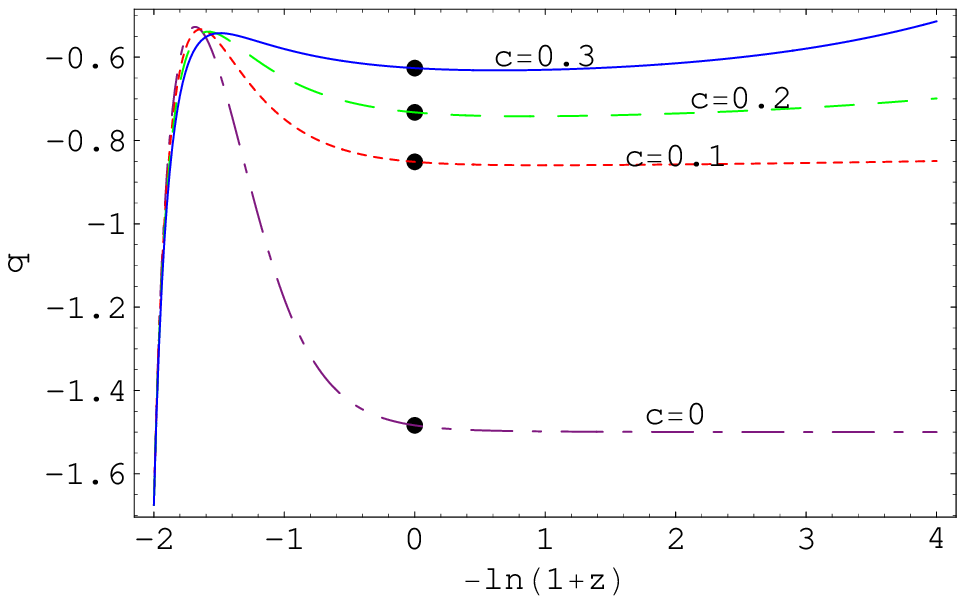}
\end{center}
\caption{The figures are the statefinder parameters versus
redshift diagram of the rolling-down scaling solution with
$x_0>0$. The left figure is $-\ln(1+z)$ --- $r$ diagram in the
variable interval $N\in[-2,20]$. The middle figure is $-\ln(1+z)$
--- $s$ diagram in the variable interval $N\in[-2,20]$. The right
figure is $-\ln(1+z)$ --- $q$ diagram in the variable interval
$N\in[-2,4]$. Selected curves for $\lambda=1$, $c=0$, $0.1$, $0.2$
and $0.3$ respectively. Dots locate the current values of the
statefinder parameters.}\label{nr2}
\end{figure}

\begin{figure}
\begin{center}
\includegraphics[width=2.4in]{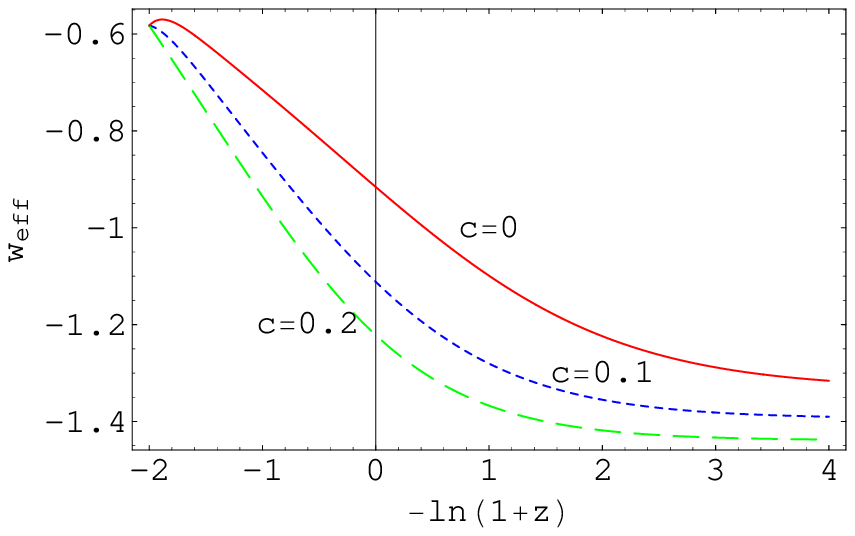}\includegraphics[width=2.4in]{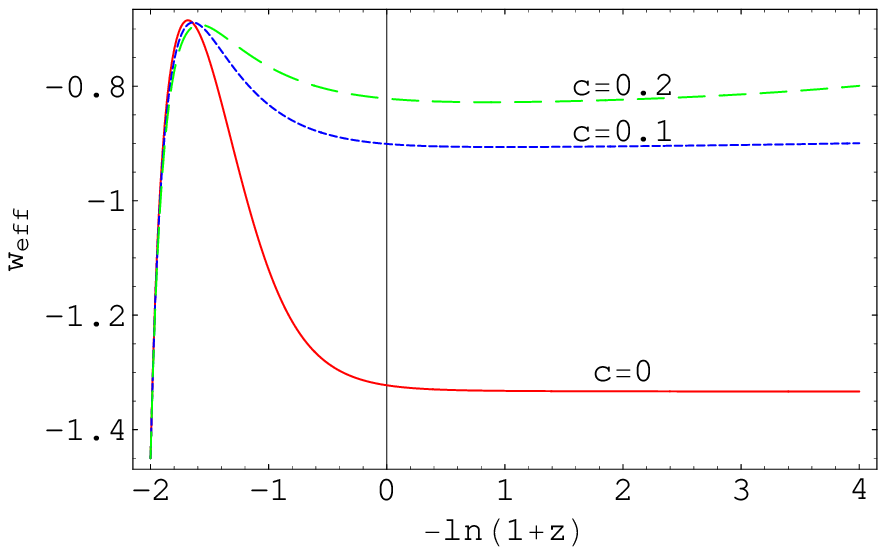}
\end{center}
\caption{The figures are the effective equation of state for the
total cosmic fluid $w_{eff}$ versus redshift. The left figure is
$-\ln(1+z)$ --- $w_{eff}$ diagram of the climbing-up scaling
solution with $x_0<0$. The right figure is $-\ln(1+z)$ ---
$w_{eff}$ diagram of the rolling-down scaling solution with
$x_0>0$. The curves evolve in the variable interval $N\in[-2,4]$.
Selected curves for $\lambda=1$, $c=0$, $0.1$ and $0.2$
respectively.}\label{w}
\end{figure}

Next we discuss another scaling solution $C$ in the case of the
interaction form (\ref{Q}) in \cite{0412624}, which exists for
$x_C>0$ ($x_C$ is also the initially velocity of the phantom field)
and
$0<{c}\leq{min\{f(\frac{-\lambda+\sqrt{\lambda^2+6}}{\sqrt{6}}),f(\frac{-\lambda+\sqrt{\lambda^2+12}}{2\sqrt{6}})\}}$
, which corresponds to a rolling-down phantom field where universe
avoids the big rip. In Fig. \ref{sr2} we show the time evolution of
the statefinder pairs $\{r,s\}$ and $\{r,q\}$ for the rolling-down
scaling solution. The plot is also for variable interval
$N\in[-2,20]$, and the corresponding parameters are also
$\lambda=1$, $c=0$, $c=0.1$, $0.2$ and $0.3$ respectively. We can
see that the trajectories of this case will pass through LCDM fixed
point. The statefinder pair $\{r,s\}$ lie in the region $s>0$, $r<1$
which is different from that in Fig. \ref{sr1}. In Fig. \ref{nr2} we
plot the evolution trajectories of the statefinder parameters versus
redshift diagram of the rolling-down scaling solution with $x_0>0$.
In the diagrams of the $-ln(1+z)$
--- $r$ and $-ln(1+z)$ --- $s$, we can see clearly that the interaction
between phantom field and dark matter causes statefinder parameters
to approach the LCDM scenario in the future. In Fig. \ref{w}, the
evolution of the effective equation of state of the total cosmic
fluid with different initial velocity have been shown. In the left
figure, the phantom field initially climbs up, and the effective
equation of state $\mathcal {W}_{eff}$ tends to below $-1$ and
realizes a transition from $\mathcal {W}_{eff}>-1$ $\mathcal
{W}_{eff}<-1$, thus the universe ends with big rip. In the right
figure, the phantom field initially rolls down, the effective
equation of state $\mathcal {W}_{eff}$ tends to above $-1$ and
realizes a transition from $\mathcal {W}_{eff}<-1$ to $\mathcal
{W}_{eff}>-1$. In this case the universe avoids the big rip.

The scaling solutions $B$ and $C$ in \cite{0412624} with different
initially velocity of the phantom field cause the different
evolution of the universe. We apply a statefinder analysis to this
model, and contrast the scenario with coupling between phantom field
and dark matter and the scenario without interaction between these
two components. The difference can be found in Fig. \ref{sr1}--Fig.
\ref{nr2}: (i) The region of the statefinder pair $\{r,s\}$ is
different between Fig. \ref{sr1} and Fig. \ref{sr2}, in Fig.
\ref{sr2} $s>0$ and $r<1$ , while in Fig. \ref{sr1} $s<0$ and $r>1$,
which differs from the quintessence and phantom models. (ii) The
influence of the interaction form is different in the evolution of
the universe, which can be seen in the statefinder trajectories. In
Fig. \ref{nr1} we can see that the interaction form causes the big
deviation between statefinder parameters and the LCDM scenario. In
Fig. \ref{nr2} the interaction causes statefinder parameters to
approach the LCDM scenario in the future. In Fig. \ref{w}, the
effective equation of state $\mathcal {W}_{eff}$ realizes a
transition from $\mathcal {W}_{eff}>-1$ to $\mathcal {W}_{eff}<-1$
and the universe ends with big rip if the phantom field initially
climbs up. Contrarily, the effective equation of state transits from
$\mathcal {W}_{eff}<-1$ to $\mathcal {W}_{eff}>-1$ and the universe
avoids the big rip if the phantom field initially rolls down. So,
through the statefinder diagnostic, we not only see the influence of
the interaction form on the evolution of universe with different
initial velocity, but also contrast the difference between the
scenario with and without the interaction.

In summary, we study the statefinder parameters of the coupled
phantom model. We analyze two cases of scaling solutions---the
climbing-up scaling solution and rolling-down scaling solution,
and contrast the difference between the scenario with and without
the interaction. It is found that the evolving trajectories of
these two scaling solutions in the $s-r$ and $q-r$ plane is quite
different, and which is also different from the statefinder
diagnostic of other dark energy models. We hope that the future
high precision observation will be capable of determining these
statefinder parameters and consequently shed light on the nature
of dark energy.

\section*{Acknowledgments} One of us (B. R. Chang) is grateful to Xin Zhang for helpful discussion. This work was supported by NSF (10573003), NSF
(10647110), NBRP (2003CB716300) of P. R. China and DUT 893321.

\end{document}